\documentclass[letter]{article}
\usepackage{graphicx}
\usepackage{geometry}
\usepackage{epsfig}
\usepackage{hyperref}
\usepackage{fancyhdr}
\usepackage{color}
\usepackage{mathrsfs}
\usepackage{indentfirst}
\usepackage{amsmath}
\usepackage{amsfonts}
\usepackage{amssymb}
\usepackage{amsthm}
\usepackage[all]{xy}
\usepackage{caption}
\usepackage[T1]{fontenc}
\usepackage{textcomp}
\usepackage{lmodern}

\makeatletter
\newcommand{\citeu}[1]{$^{\mbox{\protect \scriptsize \cite{#1}}}$}

\newcommand{\Rmnum}[1]{\expandafter\@slowromancap\romannumeral #1@}

\makeatother

\begin{document}


\title{Parameter Estimation of Switched Hammerstein Systems\renewcommand{\thefootnote}{}\footnote{This study is supported by National Natural Science Foundation of China under Grants 61273193, 61120106011, 61134013, and by the National Center for Mathematics and Interdisciplinary Sciences, Chinese Academy of Sciences. The draft has been accepted for publication by Acta Mathematicae Applicatae Sinica (http://link.springer.com/journal/10255).}}

\author{Jing Zhang  \\ \small University of Chinese Academy of Sciences (CAS), Beijing 100049, P. R. China; \small \\ \small The Key Laboratory of Systems and Control, CAS, Beijing 100080, P. R. China  \\ \small Email: zhangjing410@mails.ucas.ac.cn \\\\ Han-Fu Chen  \\ \small The Key Laboratory of Systems and Control, CAS, Beijing 100080, P. R. China\\\small Email: hfchen@iss.ac.cn}
\date{\small\it }

\maketitle

\textbf{Abstract} ~This paper deals with the parameter estimation problem of the Single-Input-Single-Output (SISO) switched Hammerstein system.
Suppose that the switching law is arbitrary but can be observed
online. All subsystems are parameterized and the Recursive Least
Squares (RLS) algorithm is applied to estimate their parameters. To
overcome the difficulty caused by coupling of data from different
subsystems, the concept \textit{intrinsic switch} is introduced. Two
cases are considered: i) The input is taken to be a sequence of
independent identically
 distributed (i.i.d.) random variables when identification is the only purpose; ii) A diminishingly excited signal is
 superimposed on the control when the adaptive control law is given.
 The strong consistency of the estimates in both cases is established and a simulation example is given to verify the theoretical analysis.


%

\section{Introduction}
Because of importance in engineering applications, the
identification and control of switched systems have been active
research areas for years\citeu{Sun2005}. Concerning parameter
identification of switched systems, a survey is given in
\cite{Vidal2007}.

The switched systems can roughly be divided into two classes:
systems with an arbitrary switching mechanism and systems governed
by a constrained switching law, such as the Markovian switching
rule. In the existing literature there are many papers on Markov
Jump Systems, see, e.g., \cite{Costa2005} and the references
therein. The Markov models are also considered in \cite{cdc15,tsp} for purposes of anomaly detection. 

By using the algebraic geometry as the key tool and under the
assumption that the number of subsystems, the subsystem orders, and
the switching sequence are unknown, the author of \cite{Vidal2008}
provides an algorithm to recursively estimate the unknown parameters
of the discrete-time Switched Auto-Regressive eXogenous (SARX)
model, and gives the algorithm a convergence analysis. However, in
the convergence analysis given in \cite{Vidal2008} no unpredictable
disturbance is taken into account, despite the examples given there
are with noises. While the authors of \cite{Bako2011} tackle the
SARX model with noises; they suggest an algorithm that alternates
between data designation to submodels and parameter update, but do
not prove its convergence. Recently, in transportation community, Zhang et. al \cite{cdc16,ifac17,cdc17} leverage Least Squares (LS) methods to ensure flow conservation and estimate Origin-Destination (OD) flow demand matrices, which have been demonstrated pretty effective and efficient, thus motivating our current work to consider a recursive version of LS.

In this work, we consider parameter estimation of the
Single-Input-Single-Output (SISO) switched Hammerstein system and
assume that the switching law is arbitrary but can be observed
online. We will handle two cases:

i) In the case where identifying the system is the only concern, we
take the system input as a sequence of i.i.d. random variables. It
is assumed that the nonlinear function of each subsystem can be
expanded to a linear combination of continuous base functions.

ii) In the case where the adaptive control has been designed for the
system, we apply the diminishing excitation
technique\citeu{ChenGuo1991} to recursively estimate the unknown
parameters. In this case, we assume that the continuous base
functions, a linear combination of which the nonlinear part of each
subsystem can be expanded to, are monomials.

The rest of the paper is organized as follows.  The problem is
formulated in Section 2, and the parameter estimation algorithm is
constructed in Section 3. In  Section 4 we prove that the estimates
given by the proposed algorithm are strongly consistent, and then we
provide a simulation example in Section  5. Some concluding remarks
are given in  Section 6. Appendix at the end is used to load proof
details.

\section{Problem Formulation}
 The SISO switched Hammerstein system considered in the paper is presented in Fig. 1. It contains  a finite number of  Hammerstein subsystems, each of which consists of a static nonlinear \(G\left(  \cdot  \right)\) followed  by an ARX subsystem in cascade.

\begin{figure}[ht]
\centering
\includegraphics[height=4cm]{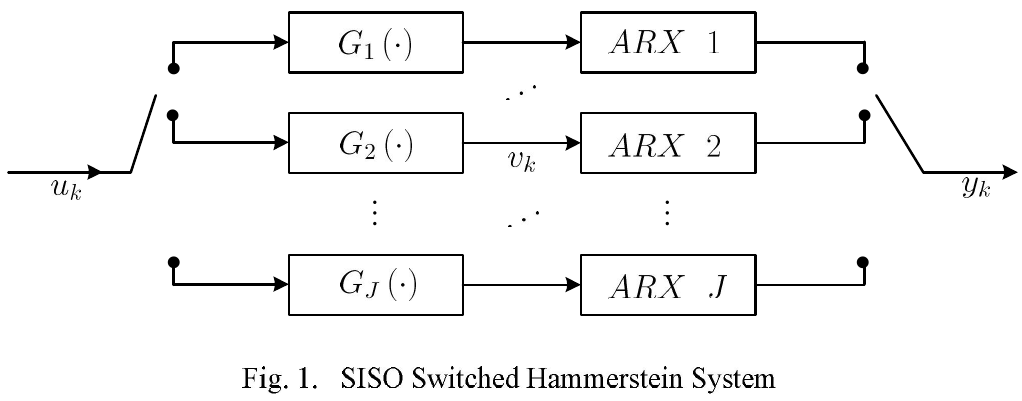}
\end{figure}

    We assume that there are \(J\) subsystems, and consider the case where the switch mechanism is available. To be precise, the mapping \(\lambda \left(  \cdot  \right)\)
    \[\begin{gathered}
    \mathbb{N}\xrightarrow{\lambda }\left\{ {1,2, \ldots ,J} \right\} \hfill \\
    ~~k \mapsto {\lambda _k} \hfill \\
    \end{gathered} \]
    can be observed online, where \(\mathbb{N}\) represents the set of all nonnegative integers, and \({\lambda _k}\) denotes the serial number of the Hammerstein subsystem that operates at time \(k\). Besides, the orders \(p,q\) of all ARX subsystems are supposed to be the same and known. Moreover, ${G_j}\left(  \cdot  \right)$, $\forall j \in \{1, \ldots ,J\}$, can be expressed as a linear combination of $r$ basis functions: ${g_1}\left(  \cdot  \right), \ldots, {g_r}\left(  \cdot  \right)$.

    By setting
\begin{align}
{A_{{\lambda _k}}}\left( z \right) \buildrel \Delta \over =& 1 + {a_1^{\scriptscriptstyle{(\lambda _k)}}}z +  \cdots  + {a_p^{\scriptscriptstyle{(\lambda _k)}}}{z^p},\nonumber\\
{B_{{\lambda _k}}}\left( z \right) \buildrel \Delta \over =& {b_1^{\scriptscriptstyle{(\lambda _k)}}} + {b_2^{\scriptscriptstyle{(\lambda _k)}}}z +  \cdots  + {b_q^{\scriptscriptstyle{(\lambda _k)}}}{z^{q-1}},\nonumber\\
{G_{{\lambda _k}}}\left(  \cdot \right) \buildrel \Delta \over =& \sum\limits_{l = 1}^r {c{}_l^{\scriptscriptstyle{(\lambda _k)}}{g_l}\left(  \cdot  \right)} , \nonumber
\end{align}
the system can be described as
\begin{align}
\left\{ \begin{gathered}
  {v_k} = {G_{{\lambda _k}}}\left( {{u_k}} \right), \hfill\\
  {A_{{\lambda _k}}}\left( z \right){y_{k + 1}} = {B_{{\lambda _k}}}\left( z \right){v_k} + {\xi _{k + 1}},~ k \geq 0; \hfill\\
  {u_k} \triangleq 0,{v_k} \triangleq 0,{\xi _{k + 1}} \triangleq 0,{y_{k + 1}} \triangleq 0,~ k < 0,
\end{gathered}  \right.
\label{eq1}
\end{align}
    where \({{u_k}}\) is the input, \({{v_k}}\) is the unmeasurable internal signal generated by \({G_{{\lambda _k}}}\left(  \cdot  \right)\), \({{y_k}}\) is the output, \({{\xi_k}}\) is the driven noise, and \(z\) denotes the backward shift operator, $zy_k=y_{k-1}$.

On the other hand, we set
\[{\tilde{A}^{\scriptscriptstyle{\left( {{k}} \right)}}} \buildrel \Delta \over = {\left[ {\begin{array}{*{20}{c}}
{ - a_1^{\scriptscriptstyle{\left( {{\lambda _k}} \right)}}}&1& \cdots &0\\
 \vdots & 0 & \ddots  & \vdots &  \\
 \vdots & \vdots &   \ddots & 1  \\
{ - a_h^{\scriptscriptstyle{\left( {{\lambda _{k + h - 1}}} \right)}}}&0& \cdots &0
\end{array}} \right]_{h \times h}},\]

\[{\tilde{B}^{\scriptscriptstyle{\left( {{k}} \right)}}} \buildrel \Delta \over = {\left[ {\begin{array}{*{20}{c}}
{b_1^{\scriptscriptstyle{\left( {{\lambda _k}} \right)}}c_1^{\scriptscriptstyle{\left( {{\lambda _k}} \right)}}}& \cdots &{b_1^{\scriptscriptstyle{\left( {{\lambda _k}} \right)}}c_r^{\scriptscriptstyle{\left( {{\lambda _k}} \right)}}}\\
 \vdots & {}  & \vdots \\
{b_h^{\scriptscriptstyle{\left( {{\lambda _{k + h - 1}}} \right)}}c_1^{\scriptscriptstyle{\left( {{\lambda _{k + h - 1}}} \right)}}}& \cdots &{b_h^{\scriptscriptstyle{\left( {{\lambda _{k + h - 1}}} \right)}}c_r^{\scriptscriptstyle{\left( {{\lambda _{k + h - 1}}} \right)}}}
\end{array}} \right]_{h \times r}},\]
$C^\tau \triangleq {\left[ {1~0 ~\cdots~ 0} \right]_{1 \times h}}$, and ${{\tilde u}^\tau_k} \triangleq {\left[ {{g_1}\left( {{u_k}} \right) ~\cdots~ {g_r}\left( {{u_k}} \right)} \right]_{1 \times r}}$,
where $h \buildrel \Delta \over = \max \{ p,q\}$, $a_l^{\scriptscriptstyle{\left( j \right)}} \buildrel \Delta \over = 0$, $b_m^{\scriptscriptstyle{\left( j \right)}} \buildrel \Delta \over = 0$,
for $l > p,m > q,j \in \left\{ {1, \ldots ,J} \right\}$. Then System \eqref{eq1} can be expressed in the state space form as follows:
\begin{align}
\left\{ \begin{gathered}
  {x_{k + 1}} = {\tilde{A}^{\scriptscriptstyle{\left( {{k}} \right)}}}{x_k} + {\tilde{B}^{\scriptscriptstyle{\left( {{k}} \right)}}}{\tilde{u}_k} + C{\xi _{k + 1}}, \hfill \\
  {y_k} = {C^\tau }{x_k}, \hfill \\
  x_0^\tau  = {\left[ {{y_0}~~0 ~\cdots~ 0} \right]_{1 \times h}} = {\left[ {0~~0 ~\cdots~ 0} \right]_{1 \times h}}. \hfill \\
\end{gathered}  \right.     \label{eq999}
\end{align}

{\bf Remark 1}~~{\it It is seen that ${\tilde{A}^{\scriptscriptstyle{\left( k \right)}}}$ and ${\tilde{B}^{\scriptscriptstyle{\left( k \right)}}}$  take values in the finite sets, which will be denoted by $\left\{ {{A^{\scriptscriptstyle{\scriptscriptstyle{{\scriptscriptstyle{{\left( 1 \right)}}}}}}}, \ldots, {A^{\scriptscriptstyle{\left( {{S_1}} \right)}}}} \right\}$ and $\left\{ {{B^{\scriptscriptstyle{\scriptscriptstyle{{\scriptscriptstyle{{\left( 1 \right)}}}}}}}, \ldots, {B^{\scriptscriptstyle{\left( {{S_2}} \right)}}}} \right\}$, respectively.}

We make the following assumption on the system.

    (H0)~For each \(j \in \left\{ {1,2, \ldots ,J} \right\}\),  \({\lambda ^{ - 1}}\left( {\left\{ j \right\}} \right)\) is an infinite subsequence of \(\mathbb{N}\), and
    \[\begin{gathered}
    {\lambda ^{ - 1}}\left( {\left\{ j_1 \right\}} \right) \cap {\lambda ^{ - 1}}\left( {\left\{j_2 \right\}} \right) = \emptyset ,{\text{ }}\forall {\text{ 1}} \leq j_1 \ne j_2 \leq J, \hfill \\
    \bigcup\nolimits_{j = 1}^J {{\lambda ^{ - 1}}\left( {\left\{ j \right\}} \right)}  = \mathbb{N}. \hfill \\
    \end{gathered} \]

{\bf Remark 2}~~{\it By (H0) we preclude those subsystems that only operate for a finite number of times; this is reasonable when processing parameter identification task.}

For System \eqref{eq1}, the parameter estimation problem is to
recursively estimate the unknown parameters $
a_1^{\scriptscriptstyle{(j)}}, \ldots,
a_p^{\scriptscriptstyle{(j)}}, ~b_1^{\scriptscriptstyle{(j)}},
\ldots, b_q^{\scriptscriptstyle{(j)}},
~c_1^{\scriptscriptstyle{(j)}}, \ldots,
c_r^{\scriptscriptstyle{(j)}}$, $\forall j \in \{1, \ldots ,J\}$,
based on the designed input \(\left\{ {{u_k}} \right\}_{k =
0}^\infty \) and the measured output  \(\left\{ {{y_k}} \right\}_{k
= 1}^\infty \).

\section{Estimation Algorithm}
Let \(j \in \left\{ {1,2, \ldots ,J} \right\}\) be arbitrarily fixed. By (H0) we are able to  write
${\lambda ^{ - 1}}\left( {\left\{ j \right\}} \right) = \left\{ {{k_{t}^{\scriptscriptstyle{(j)}}}} \right\}_{t = 0}^\infty $
with \({k_l^{\scriptscriptstyle{(j)}}} < {k_s^{\scriptscriptstyle{(j)}}}\) whenever \(0 \leq l < s\). Clearly $\left\{ {k_t^{\scriptscriptstyle{(j)}}} \right\}_{t = 0}^\infty $ denotes all the times at which the $j$th Hammerstein subsystem operates; we have \({k_t^{\scriptscriptstyle{(j)}}}\xrightarrow[{t \to \infty }]{}\infty \). It is worth noting that $y_{{k_t^{\scriptscriptstyle{\left( j \right)}}}+1}$ is generated by the $j$th subsystem, while  $y_{{k_t^{\scriptscriptstyle{\left( j \right)}}}-d}$, $\forall d \in \left\{ {0, \ldots ,p - 1} \right\}$, is not necessarily the output of the $j$th subsystem.

Let us introduce a concept named \textit{intrinsic switch}.
Corresponding to $\left[ {y_{{k_t^{\scriptscriptstyle{\left( j
\right)}}}} ~\cdots~ y_{{k_t^{\scriptscriptstyle{\left( j \right)}}}
+ 1 - p}} \right]$, we set ${n^{\scriptscriptstyle{\left( j
\right)\left( t \right)}}}\triangleq \left[
{n_{0}^{\scriptscriptstyle{\left( j \right)\left( t \right)}}
~\cdots~ n_{p - 1}^{\scriptscriptstyle{\left( j \right)\left( t
\right)}}}\right]$, where $n_{d}^{\scriptscriptstyle{\left( j
\right)\left( t \right)}}$, $d \in \left\{ {0, \ldots ,p - 1}
\right\}$ denotes  the serial number of the Hammerstein subsystem
that generates ${y_{{k_t^{\scriptscriptstyle{\left( j \right)}}} -
d}}$. It is seen that ${n^{\scriptscriptstyle{\left( j \right)\left(
t \right)}}}$ is among $J^p \triangleq K$ different combinations.
From now on, we say an \textit{intrinsic switch} occurs whenever
${n^{\scriptscriptstyle{\left( j \right)\left( t \right)}}}$
changes. Evidently, we may partition $\left\{ t \right\}_{t =
0}^\infty $ into $K$ subsequences $\left\{
{{t_m^{\scriptscriptstyle{\left( \kappa  \right)}}},m \geq 0}
\right\}$, $\kappa = 1, \ldots, K$, such that for each $\kappa  \in
\left\{ {1, \ldots ,{K}} \right\}$, ${n^{\scriptscriptstyle{\left( j
\right)\left(t_m^{\left(\kappa\right)}\right)}}}$ is independent of
$m$. It is noticed that there exists at least one $\kappa  \in
\left\{ {1, \ldots ,{K}} \right\}$ such that $\left\{
{{t_m^{\scriptscriptstyle{\left( \kappa  \right)}}},m \geq 0}
\right\}$ is an infinite subsequence of $\left\{ t \right\}_{t =
0}^\infty $.

{\bf Remark 3}~~{\it The term ``intrinsic switch'' should be
distinguished from ``switch;'' ``switch'' indicates the behavior
that System \eqref{eq1} jumps from one Hammerstein subsystem to
another.}

By the notation introduced above, we know the \(j{\text{th}}\) Hammerstein subsystem works by the following equation:
\begin{eqnarray}
\begin{gathered}
\begin{split}
  &\left( {1 + {a_1^{\scriptscriptstyle{(j)}}}z +  \cdots  + {a_p^{\scriptscriptstyle{(j)}}}{z^p}} \right){y_{{k_t^{\scriptscriptstyle{(j)}}} + 1}} \hfill \\
   =& \left( {{b_1^{\scriptscriptstyle{(j)}}} + {b_2^{\scriptscriptstyle{(j)}}}z +  \cdots  + {b_q^{\scriptscriptstyle{(j)}}}{z^{q-1}}} \right)\sum\limits_{l = 1}^r {c_l^{\scriptscriptstyle{(j)}}{g_l}({u_{{k_t^{\scriptscriptstyle{(j)}}}}})}  + {\xi _{{k_t^{\scriptscriptstyle{(j)}}} + 1}},~t = 0,1, \ldots.  \hfill \\
\end{split}
\end{gathered} \label{eq2}
\end{eqnarray}
Denoting by
\[\theta ^{\scriptscriptstyle{(j)}} \triangleq \left[ { - {a_1^{\scriptscriptstyle{(j)}}}~ \cdots  ~- {a_p^{\scriptscriptstyle{(j)}}}~~{b_1^{\scriptscriptstyle{(j)}}}{c_1^{\scriptscriptstyle{(j)}}}~ \cdots~ {b_1^{\scriptscriptstyle{(j)}}}{c_r^{\scriptscriptstyle{(j)}}}~ \cdots~ {b_q^{\scriptscriptstyle{(j)}}}{c_1^{\scriptscriptstyle{(j)}}} ~\cdots~ {b_q^{\scriptscriptstyle{(j)}}}{c_r^{\scriptscriptstyle{(j)}}}} \right]^\tau\]  and

  \[\begin{split}  \varphi _t^{\scriptscriptstyle{(j)}} \triangleq \Big[&{y_{{k_t^{\scriptscriptstyle{(j)}}}}} ~\cdots~ {y_{{k_t^{\scriptscriptstyle{(j)}}} + 1 - p}}~~{g_1}\left( {{u_{{k_t^{\scriptscriptstyle{(j)}}}}}} \right) ~\cdots~  \\
 &{g_r}\left( {{u_{{k_t^{\scriptscriptstyle{(j)}}}}}} \right)
   ~\cdots~ {g_1}\left( {{u_{{k_t^{\scriptscriptstyle{(j)}}}+1 - q}}} \right) ~\cdots~ {g_r}\left( {{u_{{k_t^{\scriptscriptstyle{(j)}}}+1 - q}}} \right)\Big]^\tau \end{split}  \]
the parameters in the \(j\)th regression subsystem and the regressor, respectively, we rewrite \eqref{eq2}  as
\begin{eqnarray}
{y_{{k_t^{\scriptscriptstyle{(j)}}} + 1}} = \theta ^{\scriptscriptstyle{(j)}\tau}{\varphi _t^{\scriptscriptstyle{(j)}}} + {\xi _{{k_t^{\scriptscriptstyle{(j)}}} + 1}}, ~t\geq0.
\end{eqnarray}

Let \({\{ {\theta }_t^{\scriptscriptstyle{(j)}}\} _{t \geq 1}}\) be
the estimates of \({\theta ^{\scriptscriptstyle{(j)}}}\). Set
\({\theta}_0^{\scriptscriptstyle{(j)}}\) arbitrarily and
\({P_0^{\scriptscriptstyle{(j)}}} \triangleq {\alpha
_0^{\scriptscriptstyle{(j)}}}I\) with some $\alpha
_0^{\scriptscriptstyle{(j)}} \in \left( {0,\frac{1}{e}} \right)$.
The RLS algorithm\citeu{ChenGuo1991} estimating \({\theta
^{\scriptscriptstyle{(j)}}}\) is defined as follows
\begin{align}
  {{ \theta }_{t + 1}^{\scriptscriptstyle{(j)}}} =& {{ \theta }_t^{\scriptscriptstyle{(j)}}} + {\tilde{a}_t^{\scriptscriptstyle{(j)}}}{P_t^{\scriptscriptstyle{(j)}}}{\varphi _t^{\scriptscriptstyle{(j)}}}({y_{{k_t^{\scriptscriptstyle{(j)}}} + 1}} - {{ \theta }^{\scriptscriptstyle{\scriptscriptstyle{(j)}}\tau}_t}{\varphi _t^{\scriptscriptstyle{(j)}}}), \label{eq_expand_5}\\
  {\tilde{a}_t^{\scriptscriptstyle{(j)}}} =& \frac{1}{{1 + \varphi _t^{\scriptscriptstyle{(j)}\tau}{P_t^{\scriptscriptstyle{(j)}}}{\varphi _t^{\scriptscriptstyle{(j)}}}}}, \label{eq3}\\
  {P_{t + 1}^{\scriptscriptstyle{(j)}}} =& {P_t^{\scriptscriptstyle{(j)}}} - {\tilde{a}_t^{\scriptscriptstyle{(j)}}}{P_t^{\scriptscriptstyle{(j)}}}{\varphi _t^{\scriptscriptstyle{(j)}}}\varphi _t^{\scriptscriptstyle{(j)}\tau}{P_t^{\scriptscriptstyle{(j)}}},  \label{eq4}\\
  \varphi _t^{\scriptscriptstyle{(j)}} =& \Big[{y_{{k_t^{\scriptscriptstyle{(j)}}}}} ~\cdots~ {y_{{k_t^{\scriptscriptstyle{(j)}}} + 1 - p}}~~{g_1}\left( {{u_{{k_t^{\scriptscriptstyle{(j)}}}}}} \right) ~\cdots~ \nonumber\\&~{g_r}\left( {{u_{{k_t^{\scriptscriptstyle{(j)}}}}}} \right) ~\cdots~ {g_1} \left( {{u_{{k_t^{\scriptscriptstyle{(j)}}}+1 - q}}} \right) ~\cdots~{g_r}\left( {{u_{{k_t^{\scriptscriptstyle{(j)}}} +1- q}}} \right)\Big]^\tau\label{eq5}.
\end{align}

By \eqref{eq3} and \eqref{eq4} it follows that \({\left( {P_{t + 1}^{\scriptscriptstyle{(j)}}} \right)^{ - 1}} = \sum\nolimits_{i = 0}^t {\varphi _i^{\scriptscriptstyle{(j)}}\varphi _i^{\scriptscriptstyle{(j)}\tau}}  + \frac{1}{{{\alpha _0^{\scriptscriptstyle{(j)}}}}}I\).

The estimates of \(b_1^{\scriptscriptstyle{(j)}}, \ldots,
b_q^{\scriptscriptstyle{(j)}},~c_1^{\scriptscriptstyle{(j)}},
\ldots, c_r^{\scriptscriptstyle{(j)}}\) can be derived from  ${\{ {{
{\theta} }_t^{\scriptscriptstyle{(j)}}}\} _{t \geq 1}}$ under some
identifiable
conditions\citeu{Zhao2010}\citeu{Bai1998}\citeu{Chaoui2005}.

\section{Convergence Analysis}
Let  \(\left( {\Omega ,\mathscr{F},P} \right)\) be the basic probability space. The following assumptions are to be used.

(H1)~\({\left\{ {{\xi _k},{\mathscr{F}_k}} \right\}}\) is a
martingale difference sequence\citeu{Chow1997}   with
  \[\mathop {\sup }\nolimits_{k} E\left[ {\left. {{{\left| {{\xi _{k + 1}}} \right|}^\beta }} \right|{\mathscr{F}_k}} \right] < \infty ~~a.s.,~~\beta  \geq 2,\]
  where \({\left\{ {{\mathscr{F}_k}} \right\}}\) is a sequence of nondecreasing sub \(\sigma\)-algebras of $\mathscr{F}$.

 (H1\textquotesingle)~\({\left\{ {{\xi _k},{\mathscr{F}_k}} \right\}}\) is a martingale difference sequence with
\[\mathop {\sup }\nolimits_{k} \left| {{\xi _k}} \right| \leq W < \infty ~~a.s.,\]
  where $W$ is a positive constant,  and \({\left\{ {{\mathscr{F}_k}} \right\}}\) is a sequence of nondecreasing sub \(\sigma\)-algebras of $\mathscr{F}$.

 (H2)~$\mathop {\lim }\nolimits_{k \to \infty } \frac{1}{k}\sum\nolimits_{i = 1}^k {\xi _i^2}  = {R_\xi } > 0~~a.s.,$ where $R_\xi$ is a constant.

 (H3)~$\left\{ {1,{g_1}\left(  \cdot  \right), \ldots ,{g_r}\left(  \cdot  \right)} \right\}$ is linearly independent over some interval $\left[ {a,b} \right]$, and  ${g_l}\left(  \cdot  \right)$, $\forall l \in \left\{ {1, \ldots ,r} \right\},$ is continuous on $\left[ {a,b} \right]$.

 (H5)~There exists a $\gamma>0$ such that  as $t \to \infty $, $\sum\limits_{i = 0}^t {y_{{k_i^{\scriptscriptstyle{(j)}}} - d}^2 = O\left( {{t^\gamma }} \right)} ~~a.s.$,  $\forall d \in \{ 0, \ldots ,$ $p - 1 \}$.

 (H5\textquotesingle)~There exists a finite positive integer $\tilde{n}$ such that $\left\| {{A^{\scriptscriptstyle{\left( {{j_1}} \right)}}}{A^{\scriptscriptstyle{\left( {{j_2}} \right)}}} \cdots {A^{\scriptscriptstyle{\left( {{j_{\tilde{n}}}} \right)}}}} \right\| < 1,~\forall {A^{\scriptscriptstyle{\left( {{j_m}} \right)}}} \in \left\{ {{A^{\scriptscriptstyle{\scriptscriptstyle{{\scriptscriptstyle{{\left( 1 \right)}}}}}}}, \ldots ,{A^{\scriptscriptstyle{\left( S_1 \right)}}}} \right\},~{\text{for }}m = 1, \ldots ,\tilde{n}$, where $\left\|  \cdot  \right\|$ is the induced $1$-norm: \[\left\| A \right\| \triangleq \mathop {\max }\nolimits_{1 \leq d_2 \leq \ell_2} \sum\nolimits_{d_1 = 1}^{\ell_1} {\left| {{a_{d_1d_2}}} \right|},~\forall A = {\left( {{a_{d_1d_2}}} \right)_{\ell_1 \times \ell_2}} \in {\mathbb{R}^{{\ell_1} \times {\ell_2}}}.\]

{\bf Remark 4}~~{\it Note that  (H5\textquotesingle), as well as
(H5), is a condition concerning stability of System \eqref{eq1}.
Stability of time-varying systems is discussed in \cite{Bauer1993}
by introducing an assumption similar to (H5\textquotesingle).}

For convenience of citation, we list a lemma here:

{\bf Lemma 1}~~{\it (Theorem 2.8 of \cite{ChenGuo1991})~~Let $\{X_k,
\mathscr{G}_k\}$ be a matrix martingale difference sequence and let
$\{M_k, \mathscr{G}_k\}$ be an adapted sequence of random matrices
with $\left\| {{M_k}} \right\| < \infty ~~a.s.,~\forall k \geq 0.$
If
  \[\mathop {\sup }\nolimits_{k} E\left[ {\left. {{{\left\| {{X_{k + 1}}} \right\|}^\alpha }} \right|{\mathscr{G}_k}} \right] < \infty ~~a.s.\]
  for some $\alpha  \in \left( {0,2} \right]$, then as $k \to \infty $
  \begin{align}
  \sum\limits_{i = 0}^k {{M_i}{X_{i + 1}}}  = O\left( {{s_k}\left( \alpha  \right){{\log }^{\frac{1}{\alpha } + \eta }}\left( {s_k^\alpha (\alpha ) + e} \right)} \right)~~a.s.,~~\forall \eta  > 0,
  \end{align}
  where ${s_k}\left( \alpha  \right) = {\left( {\sum\nolimits_{i = 0}^k {{{\left\| {{M_i}} \right\|}^\alpha }} } \right)^{\frac{1}{\alpha }}}.$}

We give the convergence analysis of Algorithm \eqref{eq_expand_5}-\eqref{eq5} for two cases as follows.

\subsection{Case \text{\Rmnum{1}}---Using the i.i.d.-Type Input}

The i.i.d.-type input is taken satisfying:

(H4)~$\{u_k\}$  is a sequence of i.i.d. random variables with
density $p\left(  \cdot  \right)$, which is positive and continuous
over $\left[ {a,b} \right]$, and vanishes outside $\left[ {a,b}
\right]$. Besides, $\{u_k\}$ is independent of $\{\xi_k\}$.

Before proving our first result (Theorem 1), we need lemmas 2-5.

{\bf Lemma 2}~~{\it (Lemma 1 of \cite{Zhao2010})~~If (H3) and (H4)
hold, then
\[\begin{gathered}
  R \triangleq E{[{g_1}\left( {{u_k}} \right) - {\mu _1} ~\cdots~ {g_r}\left( {{u_k}} \right) - {\mu _r}]^\tau } [{g_1}\left( {{u_k}} \right) - {\mu _1} ~\cdots~ {g_r}\left( {{u_k}} \right) - {\mu _r}] \hfill \\
\end{gathered} > 0,\] where ${\mu _l} \triangleq E{g_l}\left( {{u_k}} \right)$, $\forall l \in \{1, \ldots ,r\}$.}

{\bf Lemma 3}~~{\it If (H1\textquotesingle), (H3), (H4), and (H5\textquotesingle) hold, then ${y_k} = O\left( 1 \right)~a.s.,$ as $k \to \infty $.}

{\it Proof}~~The proof is straightforward since System \eqref{eq999}, and thereby System \eqref{eq1}, is a contraction mapping.    \qed

By \(\lambda _{\max }^{\scriptscriptstyle{(j)}}(t)\) and \(\lambda
_{\min }^{\scriptscriptstyle{(j)}}(t)\) we denote the largest and
smallest eigenvalue of \({\left( {P_{t +
1}^{\scriptscriptstyle{(j)}}} \right)^{ - 1}}\), respectively. The
following two lemmas are motivated by Theorems 4.1 and 6.2 in
\cite{ChenGuo1991}, respectively.

{\bf Lemma 4}~~{\it Assume that (H0) and (H1) hold, and that \({u_n}\) is \({\mathscr{F}_n}\)-measurable for all $n\geq0$. Then as \(t \to \infty \) the convergence (or divergence) rate of the estimate given by Algorithm
  \eqref{eq_expand_5}--\eqref{eq5} is expressed by
  \begin{align}
  {\left\| {\theta _{t + 1}^{\scriptscriptstyle{(j)}} - {\theta ^{\scriptscriptstyle{(j)}}}} \right\|^2} = O\left( {\frac{{\log \lambda _{\max }^{\scriptscriptstyle{(j)}}(t){{\left( {\log \log \lambda _{\max }^{\scriptscriptstyle{(j)}}(t)} \right)}^{\delta (\beta  - 2)}}}}{{\lambda _{\min }^{\scriptscriptstyle{(j)}}(t)}}} \right)~~a.s.,
  \label{eq6}
  \end{align}
  where
 $\delta\left( x \right) \triangleq \left\{ \begin{gathered}
  0,~~x \ne 0; \hfill \\
  c,~~x = 0, \hfill \\
\end{gathered}  \right.$
with arbitrary constant $c>1$.}

{\it Proof}~~Applying the same method as that used in the proof of
Theorem 4.1 in \cite{ChenGuo1991},  we arrive at the desired result.
\qed

{\bf Lemma 5}~~{\it If (H0)--(H4) hold, then the following assertions are true.

1)~It holds that
  \begin{align}\mathop {\lim \inf }\limits_{t \to \infty } \frac{{\lambda _{\min }^{\scriptscriptstyle{\left( j \right)}}\left( t \right)}}{t} > 0~~a.s. \label{eq28}
  \end{align}

2)~If, in addition, (H5) holds, then the RLS estimate given by Algorithm \eqref{eq_expand_5}--\eqref{eq5} is strongly consistent and has the following convergence rate:
\begin{align}
{{{\left\| {\theta _{t + 1}^{\scriptscriptstyle{(j)}} - {\theta ^{\scriptscriptstyle{(j)}}}} \right\|}^2} = O\left( {\frac{{\log t{{\left( {\log \log t} \right)}^{\delta (\beta  - 2)}}}}{t}} \right)}~~a.s.  \label{eq29}
\end{align}}

{\it Proof}~~Analogously to the proof of Theorem 3 in
\cite{Zhao2010}, which is motivated by the proof of Theorem 6.2 in
\cite{ChenGuo1991}, we give the detailed proof of the lemma in
Appendix. \qed

We are now in a position to give and prove our first theorem.

{\bf Theorem 1}~~{\it If (H0), (H1\textquotesingle), (H2)--(H4), and (H5\textquotesingle) hold, then
the RLS estimate given by Algorithm \eqref{eq_expand_5}--\eqref{eq5} is strongly consistent and has the following convergence rate:
\begin{align}
{\left\| {\theta _{t + 1}^{\scriptscriptstyle{\left( j \right)}} - {\theta ^{\scriptscriptstyle{\left( j \right)}}}} \right\|^2} = O\left( {{{\left( {\log t} \right)} \mathord{\left/
 {\vphantom {{\left( {\log t} \right)} t}} \right.
 \kern-\nulldelimiterspace} t}} \right)~~a.s.  \label{eq301}
 \end{align} }

{\it Proof}~~Combining Lemmas 3 and  5 yields the theorem.      \qed

\subsection{Case \text{\Rmnum{2}}---Integrating the Given Adaptive Control with a Diminishingly Excited Signal}

Assume the following assumption holds:

 (H3\textquotesingle)~${g_l}\left(  x  \right) \triangleq x^l,~\forall x \in \mathbb{R}$, $\forall l \in \{1, \ldots ,r\}$.

Let $\{\varepsilon_k\}$ be a sequence of i.i.d. random variables
with continuous distribution, and let $\{\varepsilon_k\}$ be
independent of $\{\xi_k\}$ with $E\varepsilon_k = 0$,
$E\varepsilon_k^2 = 1$, and $\left| {{\varepsilon _k}} \right| \leq
\delta_0 $, where $\delta_0 >0$ is a constant.
Define\citeu{ZhaoChen2009}
\begin{align}
v_k^{\scriptscriptstyle{\scriptscriptstyle{\scriptscriptstyle{\left( d \right)}}}} \triangleq \frac{{{\varepsilon _k}}}{{{k^{\epsilon /2}}}}    \label{eq_expand_1}
\end{align}
with $\epsilon>0$  sufficiently small such that the interval $\left( {\frac{1}{2},1 - \left( {M  + 1} \right)r\epsilon } \right]$ is nonempty, where $M = Jp + q - 1$.

Without loss of generality, we  assume ${\left\{ {\mathscr{F}_k}
\right\}}$ is rich enough such that ${\xi
_k},v_k^{\scriptscriptstyle{\scriptscriptstyle{\scriptscriptstyle{\left(
d \right)}}}} \in {\mathscr{F}_k}$. Set
$\mathscr{F}^{\textquotesingle}_{k - 1} \triangleq \sigma \left\{
{{\xi _{{i_1}}},0 \leq {i_1} \leq k,{\varepsilon _{{i_2}}},0 \leq
{i_2} \leq k - 1} \right\}$.

Motivated by Theorem 6.2 in \cite{ChenGuo1991}, we introduce the
following hypothesis.

 (H4\textquotesingle)~The given adaptive control $u_k^{\scriptscriptstyle{\scriptscriptstyle{\left( c \right)}}}$ is $\mathscr{F}_{k - 1}^{\textquotesingle}$-measurable, i.e., $u_k^{\scriptscriptstyle{\scriptscriptstyle{\left( c \right)}}} \in
\mathscr{F}_{k - 1}^{\textquotesingle}, ~\forall k$, and
$u_k^{\scriptscriptstyle{\scriptscriptstyle{\left( c \right)}}} =
O\left( 1 \right)~a.s.$, as
$k\rightarrow\infty$.

 The diminishing excitation
technique\citeu{ChenGuo1991} suggests to take
\begin{align}
{u_k} \triangleq u_k^{\scriptscriptstyle{\scriptscriptstyle{\left( c
\right)}}} + v_k^{\scriptscriptstyle{\scriptscriptstyle{\left( d
\right)}}}    \label{eq_expand_2}
\end{align}
as the actual input, where
$v_k^{\scriptscriptstyle{\scriptscriptstyle{\left( d \right)}}}$ is
given by \eqref{eq_expand_1}.

Define\citeu{ZhaoChen2009}
\[U\left( k \right) \triangleq {\left[ {\begin{array}{*{20}{c}}
  1&{}&{}&{} \\
  {C_2^1u_k^{\scriptscriptstyle{\left( c \right)}}}&1&{}&{} \\
   \vdots &{}& \ddots &{} \\
  {C_r^{r - 1}{{\left( {u_k^{\scriptscriptstyle{\left( c \right)}}} \right)}^{r - 1}}}&{C_r^{r - 2}{{\left( {u_k^{\scriptscriptstyle{\left( c \right)}}} \right)}^{r - 2}}}& \cdots &1
\end{array}} \right]_{r \times r}}{\text{and }}{\overline v_k} \triangleq {\left[ {\begin{array}{*{20}{c}}
  {v_k^{\scriptscriptstyle{\left( d \right)}}} \\
  {{{\left( {v_k^{\scriptscriptstyle{\left( d \right)}}} \right)}^2}} \\
   \vdots  \\
  {{{\left( {v_k^{\scriptscriptstyle{\left( d \right)}}} \right)}^r}}
\end{array}} \right]_{r \times 1}}.\]

The following lemma is a corollary of Lemma 4 in
\cite{ZhaoChen2009}.

{\bf Lemma 6}~~{\it  Let $\left\{ {{k_s}} \right\}_{s = 0}^\infty $ be an infinite subsequence of $\left\{ k \right\}_{k = 0}^\infty $ and let $\left\{ {{t_n}} \right\}_{n = 0}^\infty $ be an infinite subsequence of $\left\{ t \right\}_{t = 0}^\infty $. If (H4\textquotesingle) holds, then we have
\begin{align}
\frac{1}{{{t_n^{\scriptscriptstyle{1 - r\epsilon }}}}}\sum\limits_{s = 0}^{t_n} {U\left( k_s \right)\left( {{{\overline v }_{k_s}} - E{{\overline v }_{k_s}}} \right){{\left( {{{\overline v }_{k_s}} - E{{\overline v }_{k_s}}} \right)}^\tau }{U^\tau }\left( {k_s} \right)}  \geq {\tilde{c}_0}I ~~a.s.   \label{eq_expand_6}
 \end{align}
for all large enough $t_n$, where $\tilde{c}_0> 0$ may depend on sample paths.}

{\it Proof} ~~Noticing (H4\textquotesingle), we obtain \eqref{eq_expand_6} by investigating its counterpart in \cite{ZhaoChen2009} with $s$ replaced by $r$ and $\delta$ set as $0$.      \qed

Modified from Theorem 2 in \cite{ZhaoChen2009}, we have the
following theorem in parallel to Theorem 1.

{\bf Theorem 2}~~{\it If (H0), (H1\textquotesingle), (H2), and (H3\textquotesingle)--(H5\textquotesingle) hold,
then the RLS estimate given by Algorithm \eqref{eq_expand_5}--\eqref{eq5} is strongly consistent and has the following convergence rate:
\begin{align}
{\left\| {\theta _{t + 1}^{\scriptscriptstyle{\left( j \right)}} - {\theta ^{\scriptscriptstyle{\left( j \right)}}}} \right\|^2} = O\left( {\frac{{\log t}}{{{t^\alpha }}}} \right)~~a.s.,~~\forall \alpha  \in \left( {\frac{1}{2},1 - \left( {M + 1} \right)r\epsilon } \right].   \label{eq_expand_301}
 \end{align} }

{\it Proof (outline)} ~~Bearing a resemblance to the proof of Lemma 5 (see Appendix), for simplicity of notation, we omit the superscript
$\left(j\right)$.  Reviewing the proofs of
Lemma 5 and Theorem 1, we see that to prove the present theorem, it
suffices to show
\begin{align}
\mathop {\lim \inf }\limits_{t \to \infty }
\frac{1}{t^\alpha}{\lambda _{\min }}\left( {\sum\limits_{i = 0}^t
{{f_i}f_i^\tau } } \right) > 0~~a.s.,~~\forall \alpha  \in \left(
{\frac{1}{2},1 - \left( {M + 1} \right)r\epsilon } \right],
\label{eq10000}
\end{align}
where ${f_i} \triangleq \prod\nolimits_{s = 1}^J {{A_s}\left( z
\right)} {\varphi _i}$. Applying once again the method of reduction
to absurdity and the procedure of \textit{subsequence partitioning
and seeking} (see Remark 6 at the end of Appendix) as that used in
the proof of Lemma 5, and  reasoning similarly to the proof of
Theorem 2 in \cite{ZhaoChen2009} with Lemmas 1 and 6 used
repeatedly, we obtain the expected result.       \qed

\section{Simulation Example}
Consider the following system:
\begin{align}
\left\{ \begin{gathered}
  {y_{2t}} = 1.1{y_{2t - 1}}-0.28{y_{2t - 2}} + 0.5u_{2t - 1} - 1.5u_{2t - 1}^2 + 2u_{2t - 1}^3 \hfill \\
   ~~~~~~~~- 2(0.5{u_{2t - 2}} - 1.5u_{2t - 2}^2 + 2u_{2t - 2}^3) + {\xi _{2t}}, \hfill \\
  {y_{2t - 1}} = 0.8{y_{2t - 2}}-0.15{y_{2t - 3}} +0.4{u_{2t - 2}} + 1.6u_{2t - 2}^2 -0.8u_{2t - 2}^3\hfill \\
   ~~~~~~~~~~~- 3(0.4{u_{2t - 3}} + 1.6u_{2t - 3}^2 -0.8u_{2t - 3}^3) + {\xi _{2t - 1}}, \hfill \\
  t \geq 2. \hfill \\
\end{gathered}  \right.       \label{eq90}
\end{align}

Let us verify (H5\textquotesingle) for System \eqref{eq90} first.  It is seen  that
\[\begin{array}{l}
{A^{\scriptscriptstyle{{\scriptscriptstyle{{\left( 1 \right)}}}}}} = \left( {\begin{array}{*{20}{c}}
{1.1}&1\\
{ - 0.28}&0
\end{array}} \right),{A^{\scriptscriptstyle{{\left( 2 \right)}}}} = \left( {\begin{array}{*{20}{c}}
{1.1}&1\\
{ - 0.15}&0
\end{array}} \right),\\
{A^{\scriptscriptstyle{{\left( 3 \right)}}}} = \left( {\begin{array}{*{20}{c}}
{0.8}&1\\
{ - 0.15}&0
\end{array}} \right),{A^{\scriptscriptstyle{\left( 4 \right)}}} = \left( {\begin{array}{*{20}{c}}
{0.8}&1\\
{ - 0.28}&0
\end{array}} \right).
\end{array}\]
Using MATLAB to calculate, we find that (H5\textquotesingle) holds with $\tilde{n}=9$.

We now assign the noise $\left\{ {{\xi_k}} \right\}$ and the excitation source $\left\{ {{\varepsilon_k}} \right\}$, and set the initial values for Algorithm \eqref{eq_expand_5}--\eqref{eq5}. Let  \({\left\{ {{\xi_k}} \right\}_{k \geq 3}}\) be i.i.d.  and uniformly distributed on \([-3,3]\). Take \({\left\{ {{\varepsilon_k}} \right\}_{k \geq 1}}\) to be i.i.d.  and uniformly distributed on \([-2,2]\) and independent of $\{\xi_k\}$. Set $\theta _0^{\scriptscriptstyle{\left( 1 \right)}} = \theta _0^{\scriptscriptstyle{\left( 2 \right)}} = 0$ and $P_0^{\scriptscriptstyle{{\left( 1 \right)}}} = P_0^{\scriptscriptstyle{{\left( 2 \right)}}} = 0.2{I_8},$ where $I_8$ denotes the $8 \times 8$ identity matrix.

Two types of input are taken separately to serve the parameter estimation task:

\textbf{Case $\text{\Rmnum{1}}$}~~Set $u_k \triangleq \varepsilon_k$.   It is noticed that all the conditions (H0), (H1\textquotesingle), and (H2)--(H4) are fulfilled. Thus, by Theorem 1, the estimate given by Algorithm \eqref{eq_expand_5}--\eqref{eq5} is strongly consistent.

On the other hand, using the designed input and the collected output to execute Algorithm \eqref{eq_expand_5}--\eqref{eq5} twice, each running $2000$ steps, we obtain the recursive estimation for the parameters of System \eqref{eq90} as shown by Fig. 2.

\begin{figure}[ht]
\includegraphics[height=5.9cm]{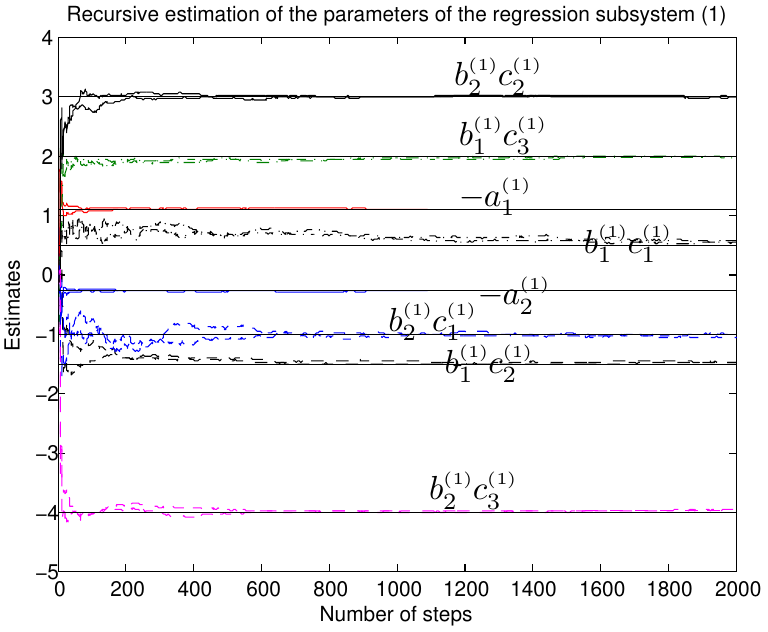}  \includegraphics[height=5.9cm]{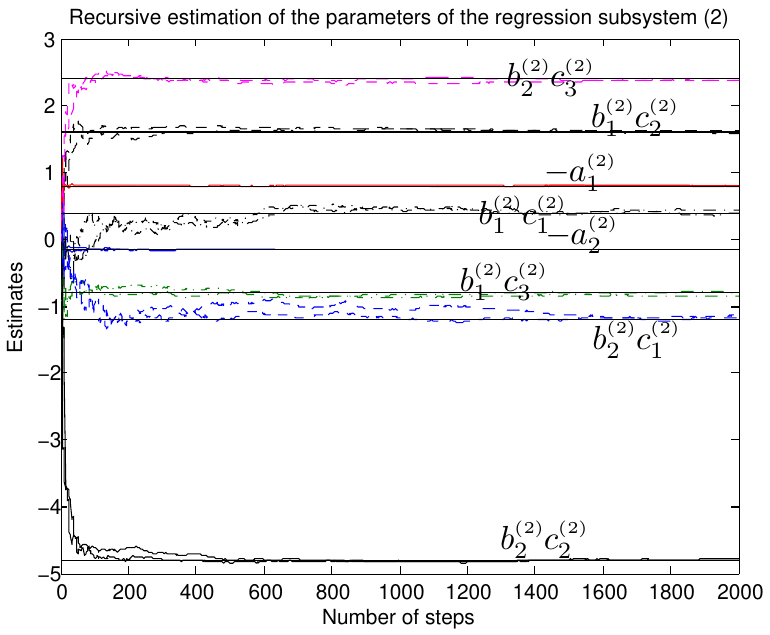}

\caption*{Fig. 2.~ Simulation results $\left(\text{\Rmnum{1}}\right)$
}
\end{figure}

\textbf{Case $\text{\Rmnum{2}}$}~~Disregarding the specific control cost, we suppose that $u_k^{\scriptscriptstyle{\left( c \right)}}
\triangleq \frac{1}{{y_k^2 + \left| {{y_{k - 1}}} \right| + 1}}$ is the given adaptive control at time $k$. Set
$v_k^{\scriptscriptstyle{\scriptscriptstyle{\left( d \right)}}}
\triangleq \frac{{{\varepsilon _k}}}{{{k^{0.001}}}}$ and ${u_k}
\triangleq u_k^{\scriptscriptstyle{\left( c \right)}} +
v_k^{\scriptscriptstyle{\scriptscriptstyle{\left( d \right)}}} =
\frac{1}{{y_k^2 + \left| {{y_{k - 1}}} \right| + 1}} +
\frac{{{\varepsilon _k}}}{{{k^{0.001}}}}$. Clearly, all the
assumptions needed by Theorem 2 hold; hence, by Theorem 2, the
estimate given by Algorithm \eqref{eq_expand_5}--\eqref{eq5} is
strongly consistent.

In this case, the corresponding simulation results are presented in Fig. 3.

\begin{figure}[ht]
\includegraphics[height=5.9cm]{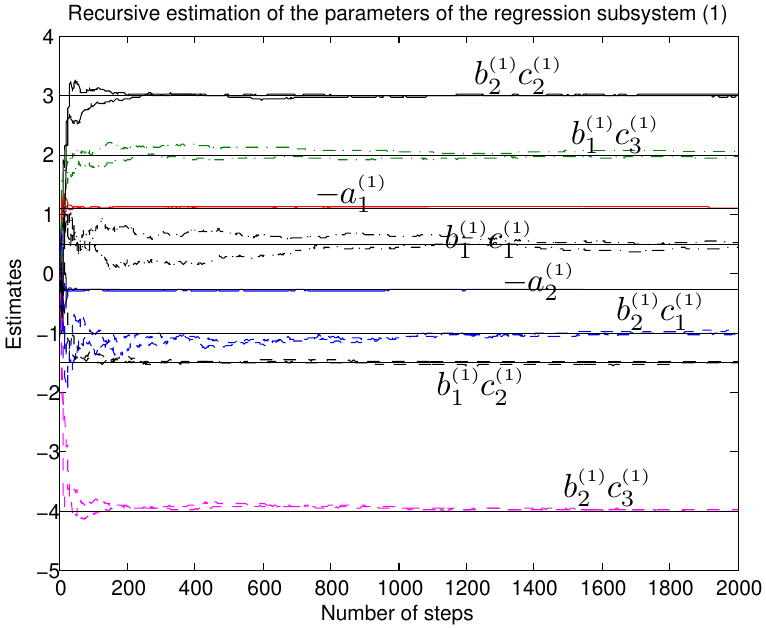}  \includegraphics[height=5.9cm]{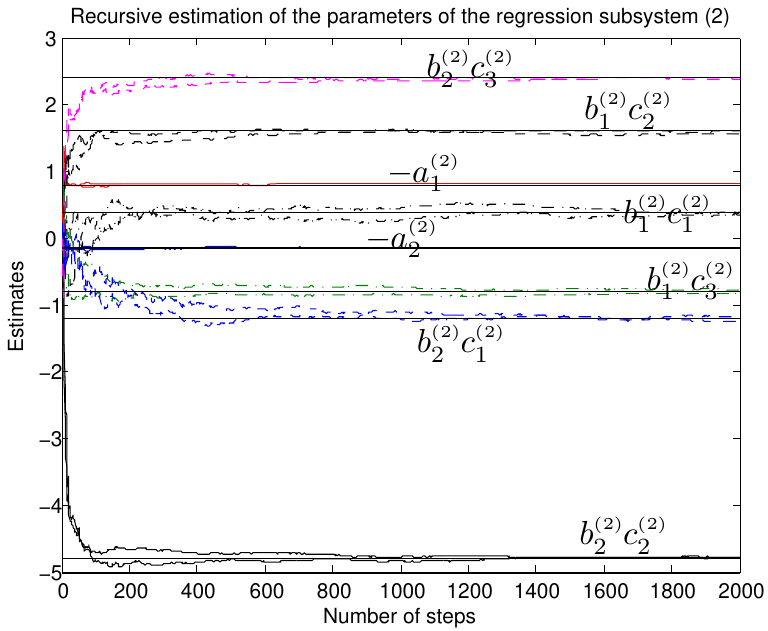}

\caption*{Fig. 3.~ Simulation results $\left(\text{\Rmnum{2}}\right)$
}
\end{figure}

It is seen that in either case the simulation outcome convincingly validates the theoretical analysis.

{\bf Remark 5}~~{\it To derive the estimates of
$b_1^{\scriptscriptstyle{{\scriptscriptstyle{{\left( 1
\right)}}}}},b_2^{\scriptscriptstyle{{\left( 1
\right)}}},c_1^{\scriptscriptstyle{{\left( 1
\right)}}},c_2^{\scriptscriptstyle{{\left( 1
\right)}}},c_3^{\scriptscriptstyle{{\left( 1
\right)}}},b_1^{\scriptscriptstyle{{\left( 2
\right)}}},b_2^{\scriptscriptstyle{{\left( 2
\right)}}},c_1^{\scriptscriptstyle{{\left( 2
\right)}}},c_2^{\scriptscriptstyle{{\left( 2
\right)}}},c_3^{\scriptscriptstyle{{\left( 2 \right)}}}$ from the
simulation results, we need to introduce appropriate identifiable
conditions, see, e.g., \cite{Zhao2010} or \cite{Bai1998} or
\cite{Chaoui2005} for details.}

\section{Concluding Remarks}
In this study, we apply the RLS algorithm to estimate the parameters
of each parameterized subsystem of the SISO switched Hammerstein
system, and under reasonable conditions we establish the strong
consistency of the estimates. Especially, in the second case, by
using the diminishing excitation technique, we also cater to
adaptive control demands. For further work, it is of interest to
consider the case where the switch mechanism is not exactly
available and to weaken the restrictions on the noise, for example,
to remove the boundedness assumption. It is also of interest to
consider the closed-loop identification problems with control costs
associated\citeu{ZhaoChen2009}.

\section{Appendix}
{\it Proof of Lemma 5}~~For simplicity of notation, we omit the superscript
$\left(j\right)$ wherever it is used to indicate the serial number of the chosen subsystem.

Define ${f_i} \triangleq \prod\nolimits_{s = 1}^J {{A_s}\left( z \right)} {\varphi _i}$. By expanding $\prod\nolimits_{s = 1}^J {{A_s}\left( z \right)} $ as $\prod\nolimits_{s = 1}^J {{A_s}\left( z \right)}=\sum\nolimits_{s = 0}^{Jp} {{\nu_s}{z^s}}$ with $\nu_0\triangleq 1$, we have ${f_i} = \sum\nolimits_{s = 0}^{Jp} {{\nu _s}{\varphi _{i - s}}} $
and
\begin{align}
  f_i^\tau
   =& \prod\limits_{s = 1}^J {{A_s}\left( z \right)} \big[{y_{{k_i}}} ~\cdots~ {y_{{k_i} + 1 - p}}~{g_1}\left( {{u_{{k_i}}}} \right) ~\cdots~  \nonumber\\
  &~~~~~~~~~~~~~~~{g_r}\left( {{u_{{k_i}}}} \right) ~\cdots~ {g_1}\left( {{u_{{k_i} + 1 - q}}} \right) ~\cdots~ {g_r}\left( {{u_{{k_i} + 1 - q}}} \right)\big] \nonumber\\
   =& \left[ \vphantom{\prod\limits_{s = 1}^J {{A_s}\left( z \right)} {g_1}\left( {{u_{{k_i} + 1 - q}}} \right) ~\cdots~ \prod\limits_{s = 1}^J {{A_s}\left( z \right)} {g_r}\left( {{u_{{k_i} + 1 - q}}} \right)}\frac{{\prod\nolimits_{s = 1}^J {{A_s}\left( z \right)} }}{{{A_{{n_0^{\scriptscriptstyle{(i)}}}}}\left( z \right)}}{B_{{n_0^{\scriptscriptstyle{(i)}}}}}\left( z \right)\sum\limits_{l = 1}^r {c_l^{\scriptscriptstyle{\left( {{n_0^{{(i)}}}} \right)}}{g_l}\left( {{u_{{k_i} - 1}}} \right)}  + \frac{{\prod\nolimits_{s = 1}^J {{A_s}\left( z \right)} }}{{{A_{{n_0^{\scriptscriptstyle{(i)}}}}}\left( z \right)}}{\xi _{{k_i}}} ~\cdots~  \right.\nonumber\\
  &~~\frac{{\prod\nolimits_{s = 1}^J {{A_s}\left( z \right)} }}{{{A_{{n_{p - 1}^{\scriptscriptstyle{(i)}}}}}\left( z \right)}}{B_{{n_{p - 1}^{\scriptscriptstyle{(i)}}}}}\left( z \right)\sum\limits_{l = 1}^r {c_l^{\scriptscriptstyle{\left( {{n_{p - 1}^{\scriptscriptstyle{(i)}}}} \right)}}{g_l}\left( {{u_{{k_i} - p}}} \right)}  + \frac{{\prod\nolimits_{s = 1}^J {{A_s}\left( z \right)} }}{{{A_{{n_{p - 1}^{\scriptscriptstyle{(i)}}}}}\left( z \right)}}{\xi _{{k_i} + 1 - p}} \nonumber\\
  &~~\prod\limits_{s = 1}^J {{A_s}\left( z \right)} {g_1}\left( {{u_{{k_i}}}} \right) ~\cdots~ \prod\limits_{s = 1}^J {{A_s}\left( z \right)} {g_r}\left( {{u_{{k_i}}}} \right) ~\cdots~  \nonumber\\
  &~~\left.\prod\limits_{s = 1}^J {{A_s}\left( z \right)} {g_1}\left( {{u_{{k_i} + 1 - q}}} \right) ~\cdots~ \prod\limits_{s = 1}^J {{A_s}\left( z \right)} {g_r}\left( {{u_{{k_i} + 1 - q}}} \right)\right],  \label{eq31}
\end{align}
where for each $d \in \left\{ {0, \ldots ,p - 1} \right\}$, by $n_d^{\scriptscriptstyle{(i)}}$ we denote the serial number of the Hammerstein subsystem that generates ${y_{{k_i} - d}}$. Clearly, $n_d^{\scriptscriptstyle{(i)}} \in \left\{ {1, \ldots ,J} \right\}$.

Using the Cauchy-Schwarz inequality, we see that
\begin{align}
  {\lambda _{\min }}\left( {\sum\limits_{i = 0}^t {{f_i}f_i^\tau } } \right)
   \leq& \mathop {\inf }\limits_{\left\| x \right\| = 1} \sum\limits_{i = 0}^t {\left( {1 + Jp} \right)\sum\limits_{s = 0}^{Jp} {\nu _s^2{{\left( {{x^\tau }{\varphi _{i - s}}} \right)}^2}} }  \nonumber\\
   \leq& \left( {1 + Jp} \right)\left( {\sum\limits_{s = 0}^{Jp} {\nu _s^2} } \right){\lambda _{\min }}\left( t \right).    \label{eq32}
\end{align}
Thus, in order to prove \eqref{eq28}, we need only to show that
\begin{align}
\mathop {\lim \inf }\limits_{t \to \infty } \frac{1}{t}{\lambda _{\min }}\left( {\sum\limits_{i = 0}^t {{f_i}f_i^\tau } } \right) > 0~~a.s.   \label{eq33}
\end{align}

We use the method of reduction to absurdity. If \eqref{eq33} were not true, then there would exist a measurable set $D$ such that $P\left\{ D \right\} > 0$ and
\begin{align}
{\mathop {\lim \inf }\limits_{t \to \infty } \frac{1}{t}{\lambda _{\min }}\left( {\sum\limits_{i = 0}^t {{f_i}f_i^\tau } } \right) = 0}, ~\forall \omega  \in D.  \label{eq34}
\end{align}

We arbitrarily choose $\omega_0 \in D$ and fix it. By \eqref{eq34} we know that there exist a subsequence ${\left\{ {{t_n}} \right\}_{n \geq 0}}$ of ${\left\{ t \right\}_{t \geq 0}}$  and a sequence of vectors ${\left\{ {{\eta _{{t_n}}}} \right\}_{n \geq 0}}$ with $\left\| {{\eta _{{t_n}}}} \right\| = 1$ such that on the sample path $\omega_0$ we have
\begin{align}  \mathop {\lim }\limits_{n \to \infty } \frac{1}{{{t_n}}}\sum\limits_{i = 0}^{{t_n}} {{{\left( {\eta _{{t_n}}^\tau {f_i}} \right)}^2}}  = 0.   \label{eq35}
\end{align}
Write ${{\eta _{{t_n}}}}$ as
\begin{align}
{\eta _{{t_n}} \triangleq \left[ {\alpha _{{t_n}}^{\scriptscriptstyle{\left( 0 \right)}} ~\cdots~ \alpha _{{t_n}}^{\scriptscriptstyle{\left( {p - 1} \right)}}~\beta _{{t_n}}^{\scriptscriptstyle{\left( {1,1} \right)}} ~\cdots~ \beta _{{t_n}}^{\scriptscriptstyle{\left( {1,r} \right)}} ~\cdots~ \beta _{{t_n}}^{\scriptscriptstyle{\left( {q,1} \right)}} ~\cdots~ \beta _{{t_n}}^{\scriptscriptstyle{\left( {q,r} \right)}}} \right]^\tau}.    \label{eq36}
\end{align}
The boundedness of $\left\{ {{\eta _{{t_n}}}} \right\}$ implies the existence of its convergent subsequence. We arbitrarily choose such a subsequence and
still use the same notation as $\left\{ {{\eta _{{t_n}}}} \right\}$ to denote it; accordingly, we are able to write
\begin{align}
{\eta _{{t_n}}}\xrightarrow[{n \to \infty }]{}\eta  \triangleq {\left[ {{\alpha ^{\scriptscriptstyle{\left( 0 \right)}}} ~\cdots~ {\alpha ^{\scriptscriptstyle{\left( {p - 1} \right)}}}~{\beta ^{\scriptscriptstyle{\left( {1,1} \right)}}} ~\cdots~ {\beta ^{\scriptscriptstyle{\left( {1,r} \right)}}} ~\cdots~ {\beta ^{\scriptscriptstyle{\left( {q,1} \right)}}} ~\cdots~ {\beta ^{\scriptscriptstyle{\left( {q,r} \right)}}}} \right]^\tau },   \label{eq55}
\end{align}
where $\left\| \eta  \right\| = 1$.

From \eqref{eq31} and \eqref{eq36} we obtain

\begin{align}
  \eta _{{t_n}}^\tau {f_i}
   =& \Bigg\{ \Bigg[\alpha _{{t_n}}^{\scriptscriptstyle{\left( 0 \right)}}\frac{{\prod\nolimits_{s = 1}^J {{A_s}\left( z \right)} }}{{{A_{{n_0^{\scriptscriptstyle{(i)}}}}}\left( z \right)}}{B_{{n_0^{\scriptscriptstyle{(i)}}}}}\left( z \right)zc_1^{{{\left(n_0^{\scriptscriptstyle{(i)}}\right)}}} ~\cdots~ \alpha _{{t_n}}^{\scriptscriptstyle{\left( 0 \right)}}\frac{{\prod\nolimits_{s = 1}^J {{A_s}\left( z \right)} }}{{{A_{{n_0^{\scriptscriptstyle{(i)}}}}}\left( z \right)}}{B_{{n_0^{\scriptscriptstyle{(i)}}}}}\left( z \right)zc_r^{{{\left(n_0^{\scriptscriptstyle{(i)}}\right)}}} \nonumber\\
   &~~~~~\alpha _{{t_n}}^{\scriptscriptstyle{\left( 0 \right)}}\frac{{\prod\nolimits_{s = 1}^J {{A_s}\left( z \right)} }}{{{A_{{n_0^{\scriptscriptstyle{(i)}}}}}\left( z \right)}}\Bigg] \nonumber\\
   &~+  \cdots  \nonumber   \\
  &~+ \Bigg[\alpha _{{t_n}}^{\scriptscriptstyle{\left( {p - 1} \right)}}\frac{{\prod\nolimits_{s = 1}^J {{A_s}\left( z \right)} }}{{{A_{{n_{p-1}^{\scriptscriptstyle{(i)}}}}}\left( z \right)}}{B_{{n_{p - 1}^{\scriptscriptstyle{(i)}}}}}\left( z \right){z^p}c_1^{{{\left(n_{p - 1}^{\scriptscriptstyle{(i)}}\right)}}} ~\cdots~
  \nonumber\\
  &~~~~~~~\alpha _{{t_n}}^{\left( p-1 \right)}\frac{{\prod\nolimits_{s = 1}^J {{A_s}\left( z \right)} }}{{{A_{{n_{p-1}^{\scriptscriptstyle{(i)}}}}}\left( z \right)}}{B_{{n_{p - 1}^{\scriptscriptstyle{(i)}}}}}\left( z \right){z^p}c_r^{{{\left(n_{p - 1}^{\scriptscriptstyle{(i)}}\right)}}} ~~\alpha _{{t_n}}^{\scriptscriptstyle{\left( {p - 1} \right)}}\frac{{\prod\nolimits_{s = 1}^J {{A_s}\left( z \right)} }}{{{A_{{n_{p - 1}^{\scriptscriptstyle{(i)}}}}}\left( z \right)}}z^{p-1}\Bigg] \nonumber\\
   &~+ \Bigg[\beta _{{t_n}}^{\scriptscriptstyle{\left( {1,1} \right)}}\prod\limits_{s = 1}^J {{A_s}\left( z \right)}  ~\cdots~ \beta _{{t_n}}^{\scriptscriptstyle{\left( {1,r} \right)}}\prod\limits_{s = 1}^J {{A_s}\left( z \right)} ~~0\Bigg] \nonumber\\
   &~+  \cdots  \nonumber\\
   &~+ \Bigg[\beta _{{t_n}}^{\scriptscriptstyle{\left( {q,1} \right)}}\prod\limits_{s = 1}^J {{A_s}\left( z \right)} {z^{q - 1}} ~\cdots~ \beta _{{t_n}}^{\scriptscriptstyle{\left( {q,r} \right)}}\prod\limits_{s = 1}^J {{A_s}\left( z \right)} {z^{q - 1}}~~0\Bigg]\Bigg\}  \nonumber\\
   &\cdot {\big[{g_1}\left( {{u_{{k_i}}}} \right) ~\cdots~ {g_r}\left( {{u_{{k_i}}}} \right)~~{\xi _{{k_i}}}\big]^\tau },   \label{eq37}
\end{align}
which can be rewritten as
\begin{align}
  \eta _{{t_n}}^\tau {f_i} \triangleq& \left[ {\sum\limits_{m = 0}^M {\tilde{h}_{{t_n}}^{\scriptscriptstyle{\left( {1,m} \right)(i)}}{z^m}}  ~\cdots~ \sum\limits_{m = 0}^M {\tilde{h}_{{t_n}}^{\scriptscriptstyle{\left( {r,m} \right)(i)}}{z^m}} ~~\sum\limits_{m = 0}^M {\tilde{h}_{{t_n}}^{\scriptscriptstyle{\left( {0,m} \right)(i)}}{z^m}} } \right]  \nonumber\\
   &\cdot {\left[ {{g_1}\left( {{u_{{k_i}}}} \right) ~\cdots~ {g_r}\left( {{u_{{k_i}}}} \right)~~{\xi _{{k_i}}}} \right]^\tau }, \label{eq38}
\end{align}
where $M=Jp+q-1$,
\begin{align}
  \sum\limits_{m = 0}^M {\tilde{h}_{{t_n}}^{\scriptscriptstyle{\left( {l,m} \right)(i)}}{z^m}}  =& \sum\limits_{m = 0}^{p - 1} {\alpha _{{t_n}}^{\scriptscriptstyle{\left( m \right)}}\frac{{\prod\nolimits_{s = 1}^J {{A_s}\left( z \right)} }}{{{A_{{n_m^{\scriptscriptstyle{(i)}}}}}\left( z \right)}}{B_{{n_m^{\scriptscriptstyle{(i)}}}}}\left( z \right){z^{m + 1}}c_l^{{\left(n_m^{\scriptscriptstyle{(i)}}\right)}}}  \nonumber\\
   &+ \sum\limits_{m = 0}^{q - 1} {\beta _{{t_n}}^{\scriptscriptstyle{\left( {m + 1,l} \right)}}\prod\nolimits_{s = 1}^J {{A_s}\left( z \right)} {z^m}} ,~~l = 1, \ldots ,r,   \label{eq39}
\end{align}
and
\begin{align}
\sum\limits_{m = 0}^M {\tilde{h}_{{t_n}}^{\scriptscriptstyle{\left( {0,m} \right)(i)}}{z^m}}  = \sum\limits_{m = 0}^{p - 1} {\alpha _{{t_n}}^{\scriptscriptstyle{\left( m \right)}}\frac{{\prod\nolimits_{s = 1}^J {{A_s}\left( z \right)} }}{{{A_{{n_m^{\scriptscriptstyle{(i)}}}}}\left( z \right)}}z^m}.   \label{eq40}
\end{align}

Recalling the concept \textit{intrinsic switch} introduced in
Section 3, we see that there exist $K$ subsequences $\left\{
i_s^{\scriptscriptstyle{(\kappa )}},s \geq 0\right\}, \kappa = 1,
\ldots, K$, of $\left\{ i \right\}_{i = 0}^\infty $ such that
$\left\{ {i_s^{\scriptscriptstyle{\left( {{\kappa _1}} \right)}}}, s
\geq 0 \right\} \bigcap \left\{ {i_s^{\scriptscriptstyle{\left(
{{\kappa _2}} \right)}}}, s \geq 0 \right\} = \emptyset ,~\forall 1
\leq {\kappa _1} \ne {\kappa _2} \leq {K}$,
$\bigcup\nolimits_{\kappa  = 1}^{{K}} {\left\{
{i_s^{\scriptscriptstyle{(\kappa )}},s \geq 0} \right\}} = \left\{ i
\right\}_{i = 0}^\infty,$ and $\left[n_0^{\scriptscriptstyle{\left(
{i_s^{\scriptscriptstyle{\left( \kappa  \right)}}} \right)}}
~\cdots~ n_{p-1}^{\scriptscriptstyle{\left(
{i_s^{\scriptscriptstyle{\left( \kappa  \right)}}} \right)}}\right],
\forall \kappa  \in \left\{ {1, \ldots, K} \right\},$ is independent
of $s$. Since for each $d \in \left\{ {0, \ldots ,p - 1} \right\}$,
$n_d^{\scriptscriptstyle{\left( {i_s^{\scriptscriptstyle{\left(
\kappa  \right)}}} \right)}}$ depends only on $\kappa$, let us
rewrite it as $n_d^{\scriptscriptstyle{(\kappa)}}$ from now on.
Obviously, there exists at least one  $\kappa  \in \left\{ {1,
\ldots ,{K}} \right\}$ such that $\left\{
{i_s^{\scriptscriptstyle{\left( \kappa \right)}},s \geq 0} \right\}$
is an infinite subsequence of $\left\{ i \right\}_{i = 0}^\infty $.
Without loss of generality, we may assume that for each $\kappa \in
\left\{ {1, \ldots ,{K}} \right\}$, $\left\{
{i_s^{\scriptscriptstyle{\left( \kappa \right)}},s \geq 0} \right\}$
is an infinite subsequence of $\left\{ i \right\}_{i = 0}^\infty $.

Let us rewrite \eqref{eq38} as
\begin{align}
  \eta _{{t_n}}^\tau {f_{i_s^{(\kappa )}}} \triangleq &\left[ {\sum\limits_{m = 0}^M {\tilde{h}_{{t_n}}^{\scriptscriptstyle{\left( {1,m} \right)\left(i_s^{(\kappa)}\right)}}{z^m}} ~\cdots~ \sum\limits_{m = 0}^M {\tilde{h}_{{t_n}}^{\scriptscriptstyle{\left( {r,m} \right)\left(i_s^{(\kappa )}\right)}}{z^m}} ~\sum\limits_{m = 0}^M {\tilde{h}_{{t_n}}^{\scriptscriptstyle{\left( {0,m} \right)\left(i_s^{(\kappa )}\right)}}{z^m}} } \right] \nonumber\\
  &~\cdot{\left[ {{g_1}\left( {{u_{{k_{i_s^{\scriptscriptstyle{(\kappa )}}}}}}} \right) ~\cdots~ {g_r}\left( {{u_{{k_{i_s^{\scriptscriptstyle{(\kappa )}}}}}}} \right)~~{\xi _{{k_{i_s^{\scriptscriptstyle{(\kappa )}}}}}}} \right]^\tau }, ~\kappa  = 1, \ldots, K,  \nonumber
\end{align}
or, equivalently,
\begin{align}
  \eta _{{t_n}}^\tau {f_{i_s^{(\kappa )}}} \triangleq &\left[ {\sum\limits_{m = 0}^M {h_{{t_n}}^{\scriptscriptstyle{\left( {1,m} \right)(\kappa )}}{z^m}}  ~\cdots~ \sum\limits_{m = 0}^M {h_{{t_n}}^{\scriptscriptstyle{\left( {r,m} \right)(\kappa )}}{z^m}} ~~\sum\limits_{m = 0}^M {h_{{t_n}}^{\scriptscriptstyle{\left( {0,m} \right)(\kappa )}}{z^m}} } \right] \nonumber\\
  &~\cdot{\left[ {{g_1}\left( {{u_{{k_{i_s^{\scriptscriptstyle{(\kappa )}}}}}}} \right) ~\cdots~ {g_r}\left( {{u_{{k_{i_s^{\scriptscriptstyle{(\kappa )}}}}}}} \right)~~{\xi _{{k_{i_s^{\scriptscriptstyle{(\kappa )}}}}}}} \right]^\tau }, ~\kappa  = 1, \ldots, K,  \label{eq63}
\end{align}
by noticing that $\tilde{h}_{{t_n}}^{\scriptscriptstyle{\left( {l,m}
\right)\left(i_s^{(\kappa)}\right)}}, ~\forall l \in \{0, 1, \ldots,
r\}$, is independent of $s$.

Corresponding to \eqref{eq39} and \eqref{eq40}, we have
\begin{align}
  \sum\limits_{m = 0}^M {h_{{t_n}}^{\scriptscriptstyle{\left( {l,m} \right)(\kappa)}}{z^m}}  =& \sum\limits_{m = 0}^{p - 1} {\alpha _{{t_n}}^{\scriptscriptstyle{\left( m \right)}}\frac{{\prod\nolimits_{s = 1}^J {{A_s}\left( z \right)} }}{{{A_{{n_m^{(\kappa )}}}}\left( z \right)}}{B_{{n_m^{(\kappa )}}}}\left( z \right){z^{m + 1}}c_l^{{\left(n_m^{\scriptscriptstyle{(\kappa )}}\right)}}}  \nonumber\\
   &+ \sum\limits_{m = 0}^{q - 1} {\beta _{{t_n}}^{\scriptscriptstyle{\left( {m + 1,l} \right)}}\prod\nolimits_{s = 1}^J {{A_s}\left( z \right)} {z^m}} ,~l = 1, \ldots ,r, ~\kappa  = 1, \ldots, K,  \label{eq64}
\end{align}
and
\begin{align}
\sum\limits_{m = 0}^M {h_{{t_n}}^{\scriptscriptstyle{\left( {0,m} \right)(\kappa)}}{z^m}}  = \sum\limits_{m = 0}^{p - 1} {\alpha _{{t_n}}^{\scriptscriptstyle{\left( m \right)}}\frac{{\prod\nolimits_{s = 1}^J {{A_s}\left( z \right)} }}{{{A_{{n_m^{(\kappa)}}}}\left( z \right)}}z^m},~\kappa  = 1, \ldots, K,   \label{eq65}
\end{align}
where $h_{{t_n}}^{\scriptscriptstyle{\left( {l,m} \right)\left( \kappa  \right)}} \in \mathbb{R},~\forall l \in \left\{ {0,1, \ldots ,r} \right\},~m \in \left\{ {0, \ldots ,M} \right\},~\kappa  \in \left\{ {1, \ldots ,K} \right\}$.
It is seen that $\left\{ {h_{{t_n}}^{\scriptscriptstyle{\left( {l,m} \right)\left( \kappa  \right)}}:l \in \left\{ {0,1, \ldots ,r} \right\},m \in \left\{ {0, \ldots ,M} \right\},\kappa  \in \left\{ {1, \ldots ,K} \right\}} \right\}$ is bounded.

We now derive from \eqref{eq35} that there exist a $\kappa  \in \left\{ {1, \ldots ,{K}} \right\}$, an infinite subsequence of $\left\{ t \right\}_{t = 0}^\infty $, and
an infinite subsequence of $\left\{ t_n \right\}_{n = 0}^\infty $,
where the latter two are denoted by $\left\{ {\tilde{t}_n^{\scriptscriptstyle{\left( \kappa  \right)}}} \right\}_{n = 0}^\infty $ and $\left\{ {{{\tilde t}_n}} \right\}_{n = 0}^\infty $, respectively, such that
\begin{align}\mathop {\lim }\limits_{n \to \infty } \frac{1}{{\tilde{t}_n^{\scriptscriptstyle{\left( \kappa  \right)}}}}\sum\limits_{s = 0}^{\tilde{t}_n^{\scriptscriptstyle{\left( \kappa  \right)}}} {{{\left( {\eta _{{\tilde{t}_n}}^\tau {f_{{i_s^{\scriptscriptstyle{\left( \kappa  \right)}}}}}} \right)}^2}}  = 0.  \label{eq786}
\end{align}

Actually, it is obvious that there exist $K$ infinite subsequences of $\left\{ t \right\}_{t = 0}^\infty $, denoted by $\left\{ {t_n^{\scriptscriptstyle{\scriptscriptstyle{{\left( 1 \right)}}}}} \right\}_{n = 0}^\infty , \ldots ,$ and $\left\{ {t_n^{\scriptscriptstyle{\left( {{K}} \right)}}} \right\}_{n = 0}^\infty $, respectively, such that for each $n \in \mathbb{N}$,
it holds that
\[\left\{ {i_s^{\scriptscriptstyle{\left( {{\kappa _1}} \right)}}} \right\}_{s = 0}^{t_n^{\left( {{\kappa _1}} \right)}} \bigcap \left\{ {i_s^{\scriptscriptstyle{\left( {{\kappa _2}} \right)}}} \right\}_{s = 0}^{t_n^{\left( {{\kappa _2}} \right)}} = \emptyset ,~~\forall 1 \leq {\kappa _1} \ne {\kappa _2} \leq {K},\]
\[\bigcup\nolimits_{\kappa  = 1}^{{K}} {\left\{ {i_s^{\scriptscriptstyle{\left( \kappa  \right)}}} \right\}_{s = 0}^{t_n^{\scriptscriptstyle{\left( \kappa  \right)}}}}  = \left\{ i \right\}_{i = 0}^{{t_n}},\]
\begin{align}
\sum\limits_{i = 0}^{{t_n}} {{{\left( {\eta _{{t_n}}^\tau {f_i}} \right)}^2}}  = \sum\limits_{\kappa  = 1}^{{K}} {\sum\limits_{s = 0}^{t_n^{\scriptscriptstyle{\left( \kappa  \right)}}} {{{\left( {\eta _{{t_n}}^\tau {f_{{i_s^{\scriptscriptstyle{\left( \kappa  \right)}}}}}} \right)}^2}} },    \label{eq781}
\end{align}
and
\[{t_n} + 1 = \sum\limits_{\kappa  = 1}^{{K}} {t_n^{\scriptscriptstyle{\left( \kappa  \right)}}}  + {K}.\]
From \eqref{eq781} and \eqref{eq35} we see that
\begin{align}\frac{1}{{{t_n} + 1}}\sum\limits_{i = 0}^{{t_n}} {{{\left( {\eta _{{t_n}}^\tau {f_i}} \right)}^2}}  = \frac{1}{{\sum\nolimits_{\kappa  = 1}^{{K}} {t_n^{\scriptscriptstyle{\left( \kappa  \right)}}}  + {K}}}\sum\limits_{\kappa  = 1}^{{K}} {\sum\limits_{s = 0}^{t_n^{\scriptscriptstyle{\left( \kappa  \right)}}} {{{\left( {\eta _{{t_n}}^\tau {f_{{i_s^{\scriptscriptstyle{\left( \kappa  \right)}}}}}} \right)}^2}} } \xrightarrow[{n \to \infty }]{}0.   \label{eq782}
\end{align}

We now show \eqref{eq786}. Assume the converse: For every $\kappa  \in \left\{ {1, \ldots ,{K}} \right\}$,
\begin{align}\mathop {\lim \inf }\limits_{n \to \infty } \frac{1}{{t_n^{\scriptscriptstyle{\left( \kappa  \right)}}}}\sum\limits_{s = 0}^{t_n^{\scriptscriptstyle{\left( \kappa  \right)}}} {{{\left( {\eta _{{t_n}}^\tau {f_{{i_s^{\scriptscriptstyle{\left( \kappa  \right)}}}}}} \right)}^2}}  > 0.   \label{eq785}
\end{align}
Then there exist a positive constant ${c_0}$ and a sufficiently large positive integer $N$ such that
\begin{align}
\sum\limits_{s = 0}^{t_n^{\scriptscriptstyle{\left( \kappa  \right)}}} {{{\left( {\eta _{{t_n}}^\tau {f_{{i_s^{\scriptscriptstyle{\left( \kappa  \right)}}}}}} \right)}^2}}  \geq {c_0}t_n^{\scriptscriptstyle{\left( \kappa  \right)}},~~\forall \kappa  \in \left\{ {1, \ldots ,{K}} \right\},~~\forall n \geq {N},  \label{eq783}
\end{align}
which leads to
\[\frac{1}{{\sum\nolimits_{\kappa  = 1}^{{K}} {t_n^{\scriptscriptstyle{\left( \kappa  \right)}}}  + {K}}}\sum\limits_{\kappa  = 1}^{{K}} {\sum\limits_{s = 0}^{t_n^{\scriptscriptstyle{\left( \kappa  \right)}}} {{{\left( {\eta _{{t_n}}^\tau {f_{{i_s^{\scriptscriptstyle{\left( \kappa  \right)}}}}}} \right)}^2}} }  \geq \frac{{{c_0}\sum\nolimits_{\kappa  = 1}^{{K}} {t_n^{\scriptscriptstyle{\left( \kappa  \right)}}} }}{{\sum\nolimits_{\kappa  = 1}^{{K}} {t_n^{\scriptscriptstyle{\left( \kappa  \right)}}}  + {K}}},~~\forall n \geq N\]
or
\[\mathop {\lim \inf }\limits_{n \to \infty } \frac{1}{{\sum\nolimits_{\kappa  = 1}^{{K}} {t_n^{\scriptscriptstyle{\left( \kappa  \right)}}}  + {K}}}\sum\limits_{\kappa  = 1}^{{K}} {\sum\limits_{s = 0}^{t_n^{\scriptscriptstyle{\left( \kappa  \right)}}} {{{\left( {\eta _{{t_n}}^\tau {f_{{i_s^{\scriptscriptstyle{\left( \kappa  \right)}}}}}} \right)}^2}} }  \geq {c_0} > 0,\]
contradicting \eqref{eq782}. Thus, \eqref{eq785} is not true, and we have shown that there exists a $\kappa  \in \left\{ {1, \ldots ,{K}} \right\}$ such that
\[\mathop {\lim \inf }\limits_{n \to \infty } \frac{1}{{t_n^{\scriptscriptstyle{\left( \kappa  \right)}}}}\sum\limits_{s = 0}^{t_n^{\scriptscriptstyle{\left( \kappa  \right)}}} {{{\left( {\eta _{{t_n}}^\tau {f_{{i_s^{\scriptscriptstyle{\left( \kappa  \right)}}}}}} \right)}^2}}  = 0,\]
which implies \eqref{eq786}.

From now on, let the $\kappa $ in \eqref{eq786} be fixed. For simplicity of notation, we omit the superscript ``~$\tilde{}$~'' in \eqref{eq786} and thereafter:
\begin{align}\mathop {\lim }\limits_{n \to \infty } \frac{1}{{{t}_n^{\scriptscriptstyle{\left( \kappa  \right)}}}}\sum\limits_{s = 0}^{{t}_n^{\scriptscriptstyle{\left( \kappa  \right)}}} {{{\left( {\eta _{{{t}_n}}^\tau {f_{{i_s^{\scriptscriptstyle{\left( \kappa  \right)}}}}}} \right)}^2}}  = 0.  \label{eq66}
\end{align}

Recall Lemmas 1 and 2. Arguing similarly to the proof of Theorem 3
in \cite{Zhao2010}, which is motivated by the proof of Theorem 6.2
in \cite{ChenGuo1991}, we derive from \eqref{eq55}, \eqref{eq63},
\eqref{eq64}, \eqref{eq65}, and \eqref{eq66} that
\begin{align}
{\eta  \triangleq {{\left[ {{\alpha ^{\scriptscriptstyle{\left( 0 \right)}}} ~\cdots~ {\alpha ^{\scriptscriptstyle{\left( {p - 1} \right)}}}~{\beta ^{\scriptscriptstyle{\left( {1,1} \right)}}} ~\cdots~ {\beta ^{\scriptscriptstyle{\left( {1,r} \right)}}} ~\cdots~ {\beta ^{\scriptscriptstyle{\left( {q,1} \right)}}} ~\cdots~ {\beta ^{\scriptscriptstyle{\left( {q,r} \right)}}}} \right]}^\tau } = 0},  \label{eq61}
\end{align}
which contradicts  ${\left\| \eta  \right\| = 1}$. Thus \textit{1)} is established.

We now prove  \textit{2)}.

To this end, recalling (H4), without loss of generality, for each $n \geq 0$, we may assume ${u_n}$ is $\mathscr{F}_n$-measurable, and therefore by Lemma 4  and  \textit{1)} of Lemma 5, we need only to show there exists a $\varrho>0$ such that
\begin{align}
{\lambda _{\max }}\left( t \right) = O\left( t^{\varrho} \right)~~a.s.  \label{eq62}
\end{align}

In fact, by (H5) it follows that
\begin{align}
  {\lambda _{\max }}\left( t \right)
   \leq& \textrm{tr}\left( {\sum\limits_{i = 0}^t {{\varphi _i}\varphi _i^\tau }  + \frac{1}{{{\alpha _0}}}I} \right) \nonumber \\
   =& O\left( {\textrm{tr}\sum\limits_{i = 0}^t {{\varphi _i}\varphi _i^\tau } } \right) = O\left( {\sum\limits_{i = 0}^t {{{\left\| {{\varphi _i}} \right\|}^2}} } \right) \nonumber \\
   =& O\left( {\sum\limits_{i = 0}^t {\sum\limits_{m = 0}^{p - 1} {y_{{k_i} - m}^2}  + \sum\limits_{i = 0}^t {\sum\limits_{m = 0}^{q - 1} {\sum\limits_{l = 1}^r {{{\left( {{g_l}\left( {{u_{{k_i} - m}}} \right)} \right)}^2}} } } } } \right) \nonumber \\
   =& O\left( {O\left( {{t^\gamma }} \right) + O\left( t \right)} \right)
   = O\left( {{t^{\varrho} }} \right)~~a.s.,
\end{align}
where $\varrho  \triangleq \max \left( {\gamma ,1} \right)$; hence, \textit{2)} is true and  the proof of Lemma 5 is completed.     \qed

{\bf Remark 6}~~{\it It is observed that throughout the proof of
Lemma 5, the procedure of deriving \eqref{eq63}--\eqref{eq786},
which can be characterized as \textit{``subsequence partitioning and
seeking,''}  plays an important role; combining this procedure with
the existing techniques applied in the proofs of Theorem 6.2 in
\cite{ChenGuo1991} and Theorem 3 in \cite{Zhao2010} leads to the
desired result.}

\section{Acknowledgment}
The authors would like to thank Professor Hai-Tao Fang for helpful discussions and valuable suggestions, and Dr. Bi-Qiang Mu  for helpful discussions.

\end{document}